\documentclass[11pt,a4paper]{article}
\pdfoutput=1

\usepackage[english]{babel}
\usepackage[latin1]{inputenc}
\usepackage{amsfonts,amsbsy,bm,bbm,euscript,mathrsfs}
\usepackage{amssymb,stmaryrd,faktor,slashed}
\usepackage{xcolor}
\usepackage[tbtags]{amsmath}
\usepackage[bookmarks=true,colorlinks=true,linkcolor=black,citecolor=black,urlcolor=black,bookmarksnumbered]{hyperref}
\usepackage[nosort]{cite}

\numberwithin{equation}{section}

\makeatletter
\renewcommand\section{\@startsection {section}{1}{\z@}
{-3.5ex \@plus -1ex \@minus -.2ex}
{2.3ex \@plus.2ex}
{\normalfont\Large\bfseries}}
\renewcommand\subsection{\@startsection{subsection}{2}{\z@}
{-3.25ex\@plus -1ex \@minus -.2ex}
{1.5ex \@plus.2ex}
{\normalfont\large\bfseries}}
\makeatother

\newcommand{\arXivlink}[1]{\href{http://arXiv.org/abs/#1}{\tt arXiv:#1}}

\newcommand{\gl}{\mathfrak{gl}}

\newcommand{\su}{\mathfrak{su}}

\newcommand{\psu}{{\mathfrak{psu}}}

\newcommand{\id}{\mathbbm 1}

\newcommand{\fn}[1]{\footnote{#1}}
\newcommand{\nn}{\nonumber}
\newcommand{\p}[1]{\left(#1\right)}

\newcommand{\f}[2]{\frac{#1}{#2}}
\newcommand{\be}[1]{\begin{equation}#1\end{equation}}
\newcommand{\ale}[1]{\begin{align}#1\end{align}}
\newcommand{\comm}[1]{\left[#1\right]}
\newcommand{\acomm}[1]{\left\{#1\right\}}

\newcommand{\bb}[1]{\mathbb{#1}}

\renewcommand{\a}{\alpha}
\newcommand{\h}{\eta}
\newcommand{\xpr}{y}
\newcommand{\eqq}{=}
\def\qfac{\sqrt{i}h}
\def\sfac{\sqrt{-i}h}
\def\ses{\widetilde E}

\begin{document}

\thispagestyle{empty}
\begin{flushright}\footnotesize\ttfamily
HU-EP-42
\\DMUS--MP--15/12
\end{flushright}
\vspace{2em}

\begin{center}

{\Large\bf \vspace{0.2cm}
{The S-matrix algebra of the $\mathbf{AdS_2 \times S^2}$ superstring}}
\vspace{1.5cm}

\textrm{\large Ben Hoare,$^{a,}$\footnote{\texttt{ben.hoare@physik.hu-berlin.de}}
Antonio Pittelli,$^{b,}$\footnote{\texttt{a.pittelli@surrey.ac.uk}}
Alessandro Torrielli$^{b,}$\footnote{\texttt{a.torrielli@surrey.ac.uk}}}

\vspace{2em}

\vspace{1em}
\begingroup\itshape
${}^a$ Institut f\"{u}r Physik und IRIS Adlershof, Humboldt-Universit\"at zu Berlin,
\\ Zum Gro\ss en Windkanal 6, 12489, Berlin, Germany.
\\\vspace{1em} ${}^b$ Department of Mathematics, University of Surrey,
\\ Guildford, GU2 7XH, UK.
\par\endgroup

\end{center}

\vspace{2em}

\begin{abstract}\noindent

In this paper we find the Yangian algebra responsible for the integrability of the $AdS_2 \times S^2 \times T^6$ superstring in the planar limit. We demonstrate the symmetry of the corresponding exact S-matrix in the massive sector, including the presence of the {\it secret} symmetry. We give two alternative presentations of the Hopf algebra. The first takes the usual canonical form, which, as the relevant representations are long, leads to a Yangian representation that is not of evaluation type. After investigating the relationship between co-commutativity, evaluation representations and the shortening condition, we find an alternative realisation of the Yangian whose representation is of evaluation type. Finally we explore two limits of the S-matrix. The first is the classical $r$-matrix, where we re-discover the need for a {\it secret} symmetry also in this context. The second is the simplifying zero-coupling limit. In this limit, taking the S-matrix as a generating R-matrix for the Algebraic Bethe Ansatz, we obtain an effective model of free fermions on a periodic spin-chain. This limit should provide hints to the {\it one-loop} anomalous dimension of the mysterious superconformal quantum mechanics dual to the superstring theory in this geometry.

\end{abstract}

\newpage

\overfullrule=0pt
\parskip=2pt
\parindent=12pt
\headheight=0.0in \headsep=0.0in \topmargin=0.0in \oddsidemargin=0in

\vspace{-3cm}
\thispagestyle{empty}
\vspace{-1cm}

\tableofcontents

\setcounter{footnote}{0}

%%%%%%%%%%%%%%%%%%%%%%%%%%%%%%%%%%%%%%
\section{Introduction}\label{secint}
%%%%%%%%%%%%%%%%%%%%%%%%%%%%%%%%%%%%%%

The remarkable impact integrability has had on the solution of string theory in
the $AdS_5 \times S^5$ background \cite{rev} motivates trying to apply the same
strategy to the other less supersymmetric string backgrounds which are still
integrable \cite{sym}. One of these is indeed the $AdS_2 \times S^2 \times T^6$
background with Ramond-Ramond fluxes in Type II superstring theory, which
preserves 8 supersymmetries. One way to generate this background is by taking
the near-horizon limit of various intersecting brane configurations in Type
IIA/B supergravity, related by T-dualities in the $T^6$ directions \cite{ads2}.
The dual "field" theory \cite{adscft} is thought to be either a superconformal
quantum mechanics, or a chiral two-dimensional CFT \cite{dual}. $AdS_2$
holography, and $AdS_2 \times S^2$ in particular, is an interesting open
problem that can be approached from several distinct directions (see for
instance \cite{gen} for some recent accounts). Given the reduced
dimensionality, it would be tempting to regard it as the simplest example where
one can test the AdS/CFT duality, but instead it turns out to be one of the
most mysterious.

The $AdS_2 \times S^2$ (coset) part of the background is conveniently encoded into a
Metsaev-Tseytlin \cite{Metsaev:1998it} type action \cite{sc}, based on the quotient
$$
\frac{PSU(1,1|2)}{ SO(1,1) \times SO(2)}~.
$$
The algebra $\mathfrak{psu}(1,1|2)$ admits a $\mathbb{Z}_4$ automorphism, which
is traditionally the key to the supercoset model being classically integrable.
Indeed, this is what happens in the $AdS_5 \times S^5$ \cite{Bena:2003wd} and
$AdS_3 \times S^3 \times M^4$ \cite{Babichenko:2009dk} cases. In the $AdS_2$
case one can truncate the Green-Schwarz action \cite{Green:1984sg} to the coset
degrees of freedom, however there is no choice of $\kappa$-symmetry gauge which
decouples the coset from the remaining fermions \cite{Sorokin:2011rr}. The
integrability of the Green-Schwarz action for the full ten-dimensional
background has been shown up to quadratic order in fermions
\cite{Sorokin:2011rr,Cagnazzo:2011at,Wulff:2015mwa}.

\

In our previous paper \cite{Hoare:2014kma}, we have utilised the symmetries of
the system and its conjectured quantum integrability to determine the exact
S-matrix for the worldsheet scattering of magnon excitations, taking the
light-cone gauge-fixed \cite{lcg} $AdS_2 \times S^2 \times T^6$ action to
infinite length. This S-matrix describes the scattering above the BMN vacuum
\cite{Berenstein:2002jq}, which is a point-like string travelling at the speed
of light on a great circle of $S^2$. The light-cone gauge-fixed Lagrangian
\cite{amsw} is highly non-trivial and breaks two-dimensional Lorentz symmetry.
Only the quadratic action preserves the Lorentz group, and describes $2+2$
(bosons+fermions) massive plus $6+6$ massless modes. The massive bosonic
excitations are associated to the transverse directions in $AdS_2 \times S^2$,
while the massless ones are associated to the $T^6$ directions.

By following a procedure which has been successful in $AdS_5$ \cite{beis0,conv}
and $AdS_3$ \cite{Borsato:2013qpa,Ya3,mf}, in \cite{Hoare:2014kma} we fixed (up
to an overall factor) the S-matrix for the excitations transforming under the
$\mathfrak{psu}(1|1)^2\ltimes \mathbb{R}$ symmetry of the BMN vacuum. In order
to do that, we relaxed the level-matching condition and postulated the presence
of two central extensions, while simultaneously deforming the coproduct in the
standard fashion \cite{tor}. The resulting massive S-matrix satisfies the
Hopf-algebra crossing relation \cite{Janik:2006dc}, and is unitarity so long as
the overall factor (dressing phase) satisfies a certain constraint. We also
studied the near-BMN expansion under certain assumptions for the dressing
phase, finding consistency with the perturbative computations
\cite{amsw,swnew}.

The main difference with the $AdS_5$ and $AdS_3$ cases is that the
representations which scatter are {\it long}, and there is no shortening
condition to be interpreted as the magnon dispersion relation. Furthermore,
because of reducibility of the tensor product representation, the S-matrix
depends on an undetermined function, which we fixed by imposing the Yang-Baxter
equation. Similar features were observed in \cite{Arutyunov:2009pw} for long
representations in $AdS_5$, and in the Pohlmeyer reduction of $AdS_2 \times
S^2$ superstrings \cite{pr}. Finally, the S-matrix enjoys an accidental $U(1)$
symmetry under which only the fermionic excitations are charged, and which is
connected to the presence of $T^6$ \cite{amsw}. This $U(1)$ allows for the
existence of a pseudo-vacuum state, and could be instrumental to derive the
Bethe equations conjectured in \cite{Sorokin:2011rr} from our S-matrix.

Although it is not completely clear what representation one should adopt for
the massless modes \cite{ads2,amsw}, in \cite{Hoare:2014kma} we took the
approach advocated in \cite{BogdanLatest} for the $AdS_3$ case, and assumed
that the massless representations and the corresponding S-matrix are the {\it
zero-mass / finite $h$} limit of the corresponding massive ones (cf.
\cite{Zamol2}), at least as far as the $\mathfrak{psu}(1|1)^2\ltimes
\mathbb{R}$ building block is concerned. We obtained in this way the limiting
S-matrices for all the choices of left and right chiralities, and discovered
that there exists a canonical Yangian for the massless sector.

\

In this paper, we obtain several results on the algebraic structure of the
exact S-matrix of the system, therefore deepening our understanding of the
associated spectral problem. Our aim is to explore the Yangian symmetry for
the massive sector. The Yangian relevant to the $AdS_5$ S-matrix was found in
\cite{Beisert:2007ds}, while for $AdS_3$ it was found in \cite{Ya3} in
separate sectors, and a larger version encompassing both left and right
algebras was discovered in \cite{Regelskis:2015xxa}. The approach we will
follow is based on the RTT formulation \cite{Molev:1994rs}, which was first
applied to the $AdS_5$ S-matrix in \cite{Beisert:2014hya}, and later to the
$AdS_3$ S-matrix in \cite{Pittelli:2014ria}.

As discussed above, the crucial new feature in the case of interest is
that the representations are long. As a result the canonical realisation of the
Yangian, similar in spirit to those in \cite{Beisert:2007ds,Ya3}, results in a
representation that is not of evaluation type. This is also a feature for long
representations in the $AdS_5$ case \cite{Arutyunov:2009pw}. After
investigating this realisation, we find a new alternative realisation of the
Yangian that does lead to a representation of evaluation type. This gives more
control over the symmetry and its action on one-particle states. Indeed the
evaluation representation is the most natural physical manifestation of
Yangians in integrable scattering problems. Furthermore, this new realisation
and the original one are contained with in a larger family of realisations
originating from a symmetry of the restricted Yangian algebra.

Besides the $\mathcal Y(\su_c(2|2))$ Yangian, the S-matrix of the $AdS_5\times
S^5$ superstring admits an additional infinite tower of conserved charges,
which constitute the so called \emph{secret symmetry} of the model
\cite{Matsumoto:2007rh,Beisert:2007ty} (see \cite{secretreview} for a review).
Such symmetries are present in several other parts of the correspondence, for
example in the pure spinor sigma model \cite{Berkovits:2011kn}, scattering
amplitudes \cite{Beisert:2011pn} and Wilson loops \cite{Munkler}. Recently,
secret symmetries were also found for the $AdS_3$ superstring
\cite{Pittelli:2014ria}, providing further evidence of their universal nature
in the AdS/CFT framework.

\

We will begin in section \ref{sec2} with a brief summary of the key properties
of Yangians that will be necessary for the following exposition. In section
\ref{sec3} we use the RTT formulation to construct the Yangian algebra
underlying the integrability of the massive sector, including the {\it secret}
symmetry. Two distinct realisations of this symmetry are given, the first of
which is close in spirit to that used in the $AdS_5$ and $AdS_3$ cases
\cite{Beisert:2014hya,Pittelli:2014ria} and in general leads to a
representation not of evaluation type, while the second is a new realisation
leading to a representation that is of evaluation type. We then discuss the
issue of evaluation representations in detail, demonstrating a relation to
shortening (massless) condition.

In section \ref{sec4} we explore the strong- and weak-coupling limits. First,
we perform a study of the classical $r$-matrix, and discover that the need for
the {\it secret} symmetry, based on the residue-analysis at the simple pole in
the spectral-parameter plane, is present also in this context. Second, we
study an effective Bethe ansatz in the simplifying limit of zero coupling, in
which the problem reduces to a standard rational spin-chain in each copy of the
symmetry algebra. In this way we obtain the spectrum of free fermions on a
periodic chain. This should represent an entry-point to the leading-order (traditionally
dubbed {\it one-loop}) anomalous dimension for the composite operators of the
mysterious superconformal quantum mechanics, supposed to be dual to the
superstring theory.

%%%%%%%%%%%%%%%%%%%%%%%%%%%%%%%%%%%%%%
\section{Yangians and integrability} \label{sec2}
%%%%%%%%%%%%%%%%%%%%%%%%%%%%%%%%%%%%%%

We start by reviewing some of the underlying formalism of Yangians and their
various realisations. Here we focus on the details that will be relevant for us
when investigating the Yangian of the massive sector of the $AdS_2 \times S^2
\times T^6$ superstring.

\paragraph{Yangians: Drinfeld's First Realisation.}

If $\mathfrak g$ is a Lie superalgebra, its \emph{Yangian} $\mathcal
Y(\mathfrak g)$ can be obtained via a quantum deformation of the universal
enveloping algebra $\mathcal U(\mathfrak g_u)$, where $\mathfrak g_u$ is the
polynomial loop algebra of $\mathfrak g$ in the variable $u$. More precisely,
\be{
\label{defini}
\mathcal Y(\mathfrak g)=\bigcup_{m\in\bb N}\mathcal Y_m(\mathfrak g) \ , \qquad
\mathcal Y_m(\mathfrak g)\eqq \text{span}\;{\mathfrak J^A_{(m)}} \ ,
}
where the vector space $\mathcal Y_m(\mathfrak g)$ corresponds to the $m$-th
\emph{level} of the Yangian $\mathcal Y(\mathfrak g)$. In the so-called {\it
Drinfeld's first realisation}, the generators $\mathfrak J^I_{(m)}$ fulfil the
graded commutation relations
\be{\label{eq: dfrgeneralcommrel}
\big[\mathfrak J^A_{(m)},\mathfrak{ J}^B_{(n)}\big\}=f^{AB}_{\quad C}\; \mathfrak{ J}^C_{(m+n)}\ , \qquad m+n=0,1 \ ,
}
with $f^{AB}_{\quad C}$ being the structure constants of $\mathfrak g$.
Equations \eqref{eq: dfrgeneralcommrel}, together with an appropriate set of
\emph{Serre relations}, determine $\mathcal Y(\mathfrak g)$ uniquely. Notice
that $\mathcal Y_0(\mathfrak g)\equiv\mathfrak g$.

Yangians enjoy the translational symmetry
\be{\label{eq: yangianshift}
\mathfrak J^A_{(0)}\to\mathfrak J^A_{(0)}\ ,\qquad\mathfrak { J}^A_{(1)}\to\mathfrak {J}^A_{(1)}+\lambda\, \mathfrak J^A_{(0)} \ , \qquad \lambda\in\bb C\ ,
}
which can be clearly seen from \eqref{defini}.

A special type of representation is the \emph{evaluation representation} $\pi_E$:
\be{
\pi_E\big(\mathfrak J^A_{(m)}\big)=u^m\, \pi \big( \mathfrak J^A_{(0)}\big) \ ,
}
where $u$ is called the \emph{spectral parameter}, and $\pi$ is a chosen
representation of $\mathfrak{g}$. However, $\pi_E$ might not exist for all
representations $\pi$. One reason possible reason for this is that inserting
$\pi_E$ into the Serre relations actually imposes strong constraints back on
$\pi$ itself, which may or may not be satisfied.

\paragraph{Yangians and integrable systems.}

Quantum systems exhibiting Yangian symmetry are integrable, as a consequence of
the fact that the Yangian allows to construct a solution to the Yang-Baxter
equation which controls the inverse scattering problem (and often the S-matrix
of the excitations) \cite{Drinfeld:1985rx}. Indeed, let $R\in A\otimes A$ be a
scattering matrix,
with $A=\mathcal Y(\mathfrak g)$ being the Hopf superalgebra whose
coproduct $\Delta:A\to A\otimes A$ defines the action of the conserved charges
on two-particles states. If $A$ and $R$ are such that
\ale{
\Delta^\text{op}(a)\,R&=R\,\Delta(a)\qquad \qquad \forall a\in A \ , && \text{(quasi co-commutativity),} \nn
\\\label{cocommutatitivity}
(\Delta \otimes \mathbbmss{1}) (R)&=R_{13}R_{23},\qquad (\mathbbmss{1} \otimes \Delta)(R)=R_{13}R_{12} \ , && \text{(quasi-triangularity),}
}
where $\Delta^\text{op}\eqq (\sigma\circ\Delta)$ is the opposite coproduct,
$\sigma$ the graded permutation operator and the subscripts 1,2,3 indicate the
copy of $A$ in the triple tensor product, then $R$ obeys the \emph{quantum
Yang-Baxter equation} (QYBE)
\be{\label{eq: qybe}
R_{12}R_{13}R_{23}=R_{23}R_{13}R_{12}
}
(the hallmark of integrability) and the \emph{crossing symmetry} equations.
This is the content of a famous theorem of Drinfeld, proving the crucial role
played by quantum groups in producing solutions to the QYBE with the desired
symmetry properties \cite{Torrielli:2011gg}.

\paragraph{RTT Realisation of the Yangian.}

Given a scattering matrix $R_{12}(u,v)$ that is a function of two spectral
parameters and satisfies the QYBE, one can extract the symmetries of the system
using the RTT realisation of the corresponding Yangian
\cite{Molev:1994rs,Beisert:2014hya}:
\be{\label{eq: rttrel}
R_{12}(u,v)\,\mathcal T_{13}(u)\mathcal T_{23}(v)=\mathcal T_{23}(v)\mathcal T_{13}(u)\,R_{12}(u,v) \ .
}
The monodromy matrix $\mathcal T(u)$ plays the role of a generating function
for the Yangian charges, which are in turn symmetries of $R$. We shall now
summarise the main features of this construction: see \cite{Beisert:2014hya}
for further details. We will focus on the $\gl(n|n)$ case, the one of interest.

Let $\acomm{e^A{}_B}$ be the standard basis for the $\gl(n|n)$ Lie
superalgebra. That is the matrices $e^A{}_B$ are such that their only
non-vanishing entry is $(-)^{[B]}$ in row $A$, column $B$. The symbol $[A]$
stands for the Gra\ss mann grading of the index $A$. $\mathcal T(u)$ can then
be written as follows:
\be{\label{eq: caltdef}
\mathcal T(u)\eqq \sum_{A,B}(-)^{[B]}\,e^B{}_A\otimes \bb T^A{}_B(u)\ ,\qquad \bb T^A{}_B:\bb C\to\mathcal Y(\gl(n|n))\ .
}
Assuming that $\bb T^A{}_B(u)$ is holomorphic in a neighbourhood of $u=\infty$, the asymptotic expansion
\be{\label{eq: bbtdefs}
\bb T^A{}_B(u)=\sum_{l\in\bb N}u^{-l}\,\bb T^A_{l-1B} \ ,
}
is well defined. At this point, one finds that particular combinations of the
$\bb T^A_{l-1B}$ can be engineered to define Drinfeld's first realisation of
$\mathcal Y(\gl(n|n))$. In particular, defining
\be{\label{eq: jdefs}
\bb U^{[B]} \delta^A_B\eqq \bb T^A_{-1B} \ , \qquad \bb J^A_{0B}\eqq \bb U^{-[B]}\,\bb T^A_{0B} \ , \qquad \bb J^A_{1B}\eqq \bb U^{-[B]}\,\bb T^A_{1B}-\f12 \bb J^A_{0C}\, \bb J^C_{0B}\ ,
}
$\bb J_0$ and $\bb J_1$ span $\gl(n|n)$ and $\mathcal Y_1(\gl(n|n))$,
respectively. The central element $\bb U$ is the \emph{braiding factor}, and
represents a deformation of the co-algebra structure. $\bb U = \id$ represents
the undeformed case.

The key observation is that the R-matrix is a particular representation of
$\mathcal T$, namely, $R(u,v)=\p{\id\otimes\pi_v}\mathcal T(u)$, where $\pi_v$
indicates a representation depending on the spectral parameter $v$. Therefore,
the coefficients in the Laurent expansion of $R(u,v)$ can be understood as $\bb
T^A_{l-1B}$ for the RTT realisation of the underlying symmetry in the
representation of interest. This can then be recast in the form of Drinfeld's
first realisation using \eqref{eq: jdefs}. The (representation-independent)
graded commutation relations for the $\bb J^A_{mB}$ are obtained from those for
the $\bb T^A_{l-1B}$, by plugging \eqref{eq: caltdef} and \eqref{eq: bbtdefs}
into \eqref{eq: rttrel} and expanding with respect to both $u$
and $v$.

Finally, the complete Hopf algebra structure of $\mathcal Y(\gl(n|n))$, in
particular, the coproducts and antipodes, can be recovered from the RTT
realisation of the Yangian. First, the fusion relation
\be{\label{eq: bbtcoproducts}
\Delta(\bb T^A{}_B(u))=\bb T^A{}_C(u)\otimes \bb T^C{}_B(u)\ ,}
descending from the R-matrix fusion relations, provides the coproducts for the
individual generators via the same expansion: indeed, expanding ${{\bb
T}^A}_B(u)$ in inverse powers of $u$ gives
\ale{\label{eq: coproductsforTs}
&\Delta(\bb U)=\bb U\otimes \bb U\ ,\qquad \Delta(\bb T^A_{0B})= \bb T^A_{0B}\otimes \bb U^{[B]}+\bb U^{[A]}\otimes\bb T^A_{0B}\ ,\nn\\
&\Delta(\bb T^A_{1B})=\bb T^A_{1B}\otimes \bb U^{[B]} +\bb U^{[A]}\otimes\bb T^A_{1B}+\bb T^A_{0C}\otimes\bb T^C_{0B} \ .
}
This can be used to derive the coproducts for $\bb J_{0,1}$:
\ale{
&\Delta(\bb U) = \bb U \otimes \bb U \ , \qquad \Delta(\bb J^A_{0B}) = \bb J^A_{0B} \otimes \id + \bb U^{[A]-[B]} \otimes \bb J^A_{0B}\nn\vphantom{\f12}
\\& \Delta(\bb J^A_{1B}) = \bb J^A_{1B} \otimes \id + \bb U^{[A]-[B]} \otimes \bb J^A_{1B} \nn\\ & \qquad\qquad + \frac12 \bb U^{[C]-[B]}\bb J^A_{0C} \otimes \bb J^C_{0B}
- \frac12 (-)^{([A]+[C])([B]+[C])} \bb U^{[A]-[C]} \bb J^C_{0B} \otimes \bb J^A_{0C} \ .
\label{eq: jsgeneralcoproducts}
}
Second, the antipode $\Sigma$ is a graded linear anti-homomorphism satisfying
\be{\label{eq: bbtcrossing}
\Sigma\comm{\bb T^A{}_C(u)}\bb T^C{}_B(u)=\bb T^A{}_C(u)\Sigma\comm{\bb T^C{}_B(u)}=\delta^A_{\;B}\ ,
\qquad \Sigma\comm{XY}=(-)^{[X][Y]} \Sigma\comm Y\Sigma\comm X\ .
}
By expanding:
\ale{
&\Sigma\comm{{\bb T}^A_{-1B}}=\bb U^{-[B]}{\delta^A}_B\ , \qquad \Sigma\comm{{\bb T}^A_{0B}}=-\bb U^{-[A]-[B]}{{\bb T}^A_{0B}}\ ,\nn\\
&\Sigma\comm{{\bb T}^A_{1B}}=-\bb U^{-[A]-[B]}{{\bb T}^A_{1B}}+\bb U^{-[A]-[B]-[C]}{{\bb T}^A_{0C}}{{\bb T}^C_{0B}}\ ,\label{antipodeexpt}
}
which in turn can be used to determine the antipodes for $\bb J_{0,1}$
\ale{
&\Sigma\comm{\bb U}=\bb U^{-1}\ ,\qquad \Sigma\comm{\bb J^A_{0B}}=-\bb U^{[B]-[A]}{{\bb J}^A_{0B}}\ ,\nn\\
&\Sigma\comm{{\bb J}^A_{1B}}=-\bb U^{[B]-[A]}{\bb J}^A_{1B}+\frac12\bb U^{[B]-[A]}[{\bb J}^A_{0C},{\bb J}^C_{0B}\}\ .\label{antipodeexp}
}

%%%%%%%%%%%%%%%%%%%%%%%%%%%%%%%%%%%%%%%%
\section{RTT realisation for the $\mathcal Y(\mathfrak{gl}_c(1|1))$ Yangian} \label{sec3}
%%%%%%%%%%%%%%%%%%%%%%%%%%%%%%%%%%%%%%%%

In this section we employ the techniques of \cite{Molev:1994rs,Beisert:2014hya}
to construct the RTT realisation for the $\mathcal Y(\gl_c(1|1))$ Yangian. The
starting point is the $\su_c(1|1)$ R-matrix of \cite{Hoare:2014kma}, which
describes the scattering of one bosonic state $|\phi\rangle$ and one fermionic
state $|\psi\rangle$ transforming in a long representation of the
centrally-extended algebra $\su_c(1|1)$
\begin{equation}\label{algrel}
\acomm{\bb Q,\bb Q}=2\,\bb P\ , \qquad \acomm{\bb S,\bb S}=2\,\bb K \ , \qquad \acomm{\bb Q,\bb S}=2\,\bb H \ .
\end{equation}
The explicit form of the representation is given by
\begin{align}\nonumber
\bb Q |\phi\rangle & = a |\psi\rangle \ , &
\bb Q |\psi\rangle & = b |\phi\rangle \ , &
\bb S |\phi\rangle & = c |\psi\rangle \ , &
\bb S |\psi\rangle & = d |\phi\rangle \ , %v2
\\\label{reps}
\bb P |\Phi\rangle & = P |\Phi\rangle \ , &
\bb K |\Phi\rangle & = K |\Phi\rangle \ , &
\bb H |\Phi\rangle & = H |\Phi\rangle \ , &
|\Phi\rangle & \in \{|\phi\rangle,\,|\psi\rangle\} \ .
\end{align}
where the eigenvalues of the central elements $\bb P$, $\bb K$ and $\bb H$
are given by
\begin{equation}
P = ab \ , \qquad K = c d \ , \qquad 2H = a d + b c \ ,
\end{equation}
as a consequence of the algebra relations \eqref{algrel}. There are no further
conditions on the central elements for the algebra to close and hence this
2-dimensional representation is long.

For generic values of $P$, $K$ and $H$ the tensor product of two of these
representations gives a 4-dimensional representation that is fully reducible
into two 2-dimensional representations.\footnote{At the special (massless)
point $H_{\rm tot}^2 - P_{\rm tot} K_{\rm tot} = 0$ the tensor product is still
reducible but no longer decomposable. Here the subscript ``tot'' indicates
that these are the eigenvalues of the central charges acting on the tensor
product state.} Therefore, the R-matrix acting on the tensor product is fixed
up to two functions by demanding invariance under the symmetry
\begin{equation}
\Delta^{\text{op}}(\mathbb J)\, R = R\, \Delta (\mathbb J) \ , \qquad \bb J = \{\bb Q, \bb S, \bb P, \bb K, \bb H\} \ .
\end{equation}
Here $\Delta$ is the coproduct, while $\Delta^{\text{op}}$ is the opposite
coproduct defined in \eqref{cocommutatitivity}. As usual for integrable
systems arising in the context of the AdS/CFT correspondence the coproduct is
deformed through the introduction of an abelian generator $\bb U$
\begin{align}\nonumber
\Delta(\bb Q) & = \bb Q \otimes \id + \bb U \otimes \bb Q \ , &
\Delta(\bb S) & = \bb S \otimes \id + \bb U^{-1} \otimes \bb S \ , &
\Delta(\bb U) & = \bb U \otimes \bb U \ ,
\\\label{coprod0}
\Delta(\bb P) & = \bb P \otimes \id + \bb U^2 \otimes \bb P \ , &
\Delta(\bb K) & = \bb K \otimes \id + \bb U^{-2} \otimes \bb K \ , &
\Delta(\bb H) & = \bb H \otimes \id + \bb \id \otimes \bb H \ . &
\end{align}
To admit an R-matrix the coproducts of the central elements should be
co-commutatitve, i.e. $\Delta^{\text{op}}(\mathbb{C}) = \Delta(\mathbb{C})$.
This relates the central charges $\bb P$ and $\bb K$ to the braiding factor
$\bb U$:
\begin{equation}\label{restrict}
\bb P = \frac h2 (1-\bb U^2) \ , \qquad \bb K = \frac h2(1 - \bb U^{-2}) \ ,
\end{equation}
where without loss of generality we have taken the constants of proportionality
to be equal. In the following we will refer to the algebra \eqref{algrel} with
these relations \eqref{restrict} imposed as the restricted algebra. While the
symmetry only constrains the R-matrix up to two functions, demanding that the
QYBE is solved (along with imposing various physical requirements such as
crossing symmetry and a sensible strong coupling limit) fixes the R-matrix up
to a single overall factor.

For reference we will quote the necessary details of the $\su_c(1|1)$ R-matrix
from \cite{Hoare:2014kma}. The R-matrix in terms of the usual Zhukovsky
variables \cite{beis0,beis1}
\begin{equation} \label{xx}
\frac{x^+}{x^-} = U^2 \ , \qquad
x^+ - \frac 1{x^+} - x^- + \frac 1{x^-} = \frac{iH}{h} \ , \qquad
x^+ + \frac{1}{x^+} -x^- - \frac{1}{x^-} = \frac{iM}{h} \ .
\end{equation}
Here
\begin{equation}\label{eq:m}
M = \frac{ad-bc}2 = \sqrt{H^2 - P K} \ ,
\end{equation}
is unconstrained as the representation \eqref{reps} is long. The
representation parameters $a$, $b$, $c$ and $d$ are given in are given in terms
of the Zhukovsky variables in \cite{Hoare:2014kma} -- we set the parameter $\a$
used there, which controls the normalisation of the bosonic state relative to
the fermionic state, to one. The R-matrix is then given by
\begin{eqnarray}
&& R \left|\phi_x\phi_y\right> = S_1 \left|\phi_x\phi_y\right> + Q_1 \left|\psi_x\psi_y\right> \ , \qquad
R\left|\psi_x\psi_y\right> = S_2 \left|\psi_x\psi_y\right> + Q_2 \left|\phi_x\phi_y\right> \ ,\nonumber
\\
\label{redu}
&&R\left|\phi_x\psi_y\right> = T_1 \left|\phi_x\psi_y\right> + R_1 \left|\psi_x\phi_y\right> \ , \qquad
\label{ans}
R\left|\psi_x\phi_y\right> = T_2 \left|\psi_x\phi_y\right> + R_2 \left|\phi_x\psi_y\right> \ ,
\end{eqnarray}
with $x^\pm$ the kinematic variables associated to the first representation and
$\xpr^\pm$ to the second. The parameterising functions are given by
\cite{Hoare:2014kma}
\begin{align}\nonumber
& S_1 = \sqrt{\frac{x^+\xpr^-}{x^-\xpr^+}}\frac{x^- - \xpr^+}{x^+ - \xpr^-} \frac{1+s_1}{2} P_0 \ ,
&& S_2 = \frac{1+s_2}{2} P_0 \ ,
\\\nonumber
& T_1 =\sqrt{\frac{\xpr^-}{\xpr^+}} \frac{x^+ - \xpr^+}{x^+ - \xpr^-} \frac{1+t_1}{2} P_0\ ,
&& T_2 =\sqrt{\frac{x^+}{x^-}} \frac{x^- - \xpr^-}{x^+ - \xpr^-}\frac{1+t_2}{2} P_0 \ ,
\\
& Q_1 = Q_2 = - \frac{i}{2} \,\sqrt[4]{\frac{x^-\xpr^+}{x^+\xpr^-}}
\frac{\h_x\h_y}{x^+-\xpr^-} \frac f{x^-x'^+} P_0 \ , &&
R_1 = R_2 = -\frac{i}{2} \,
\sqrt[4]{\frac{x^+\xpr^-}{x^-\xpr^+}} \frac{\h_x\h_y}{x^+ - \xpr^-} P_0 \ ,\label{exacta} 
\end{align}
where
\begin{align}\nonumber
f = & \frac{\sqrt{\frac{x^+}{x^-}}(x^--\frac{1}{x^+}) - \sqrt{\frac{y^+}{y^-}}(y^--\frac{1}{y^+})}{1 - \frac{1}{x^+x^-y^+y^-}} P_0 \ , &
s_1 = & \frac{1- \frac{1}{x^+\xpr^-}}{x^- - \xpr^+} f \ , \qquad s_2 = \frac{1-\frac{1}{x^-\xpr^+}}{x^+ - \xpr^-} f \ ,
\\\label{exacta1}
\h_x = & \sqrt{i(x^- - x^+)} \ , \quad \h_y = \sqrt{i(y^- - y^+)} \ , &
t_1 = & \frac{1 - \frac{1}{x^-\xpr^-}}{x^+ - \xpr^+} f \ , \qquad t_2 = \frac{1 - \frac{1}{x^+\xpr^+}}{x^- - \xpr^-} f \ .
\end{align}
As discussed above, there is an overall factor $P_0$ that is not fixed by the
considerations of symmetry and as such will not be relevant for the following
analysis. For concreteness we will fix $P_0$ such that $S_1 = 1$ following \cite{Beisert:2014hya} and so that the expansion of the R-matrix takes the form outlined in section \ref{sec2}.

Finally, let us observe that the $AdS_2 \times S^2$ worldsheet S-matrix
underlying the scattering of the massive modes is built from the tensor product
of two copies of this centrally-extended $\mathfrak{su}(1|1)$ R-matrix.

%%%%%%%%%%%%%%%%%%%%%%%%%%%%%%%%%%%%%%
\subsection{The $\mathcal Y(\su_c(1|1))$ and $\mathcal Y(\gl_c(1|1))$ Yangians and their RTT realisation}
%%%%%%%%%%%%%%%%%%%%%%%%%%%%%%%%%%%%%%

The centrally extended $ \mathcal Y(\su_c(1|1))$ Yangian is defined by the
graded commutation relations
\ale{\label{eq: suc(1|1)commrels}
\acomm{\bb Q_m,\bb Q_n}=2\,\bb P_{m+n}\ ,\qquad \acomm{\bb S_m,\bb S_n}=2\,\bb K_{m+n}\ ,\qquad
\acomm{\bb Q_m,\bb S_n}=2\,\bb H_{m+n} \ ,}
which extend the algebra \eqref{algrel}. Here, $m,n \geq 0$ indicate the level
of the corresponding generator, with $\mathbb{Q}_0$, $\mathbb{S}_0$,
$\mathbb{P}_0$, $\mathbb{K}_0$ and $\mathbb{H}_0$ playing the role of the
original generators in \eqref{algrel}.

It is worth noting that, in contrast to ordinary situations, the
infinite-dimensional algebra \eqref{eq: suc(1|1)commrels} contains infinitely
many finite-dimensional $\su_c(1|1)$ subalgebras. Indeed, the set of generators
$\mathbb{Q}_{\hat m}$, $\mathbb{S}_{\hat n}$, $\mathbb{P}_{2 \hat m}$,
$\mathbb{K}_{2\hat n}$ and $\mathbb{H}_{\hat m + \hat n}$ forms a subalgebra
for all $\hat m,\hat n\geq 0$. This is due to the fact that the generators on
the right-hand side of the relations \eqref{eq: suc(1|1)commrels} are central.

Another consequence of this, again in contrast to usual, is that for an
arbitrary representation one cannot generate the whole infinite dimensional
algebra by (anti) commuting a finite set of generators. However, considering
the natural lift of the 2-dimensional representation \eqref{reps}
\begin{align}\nonumber
\bb Q_m |\phi\rangle & = a_m |\psi\rangle \ , &
\bb Q_m |\psi\rangle & = b_m |\phi\rangle \ , &
\bb S_m |\phi\rangle & = c_m |\psi\rangle \ , &
\bb S_m |\psi\rangle & = d_m |\phi\rangle \ ,
\\\label{repsyang}
\bb P_m |\Phi\rangle & = P_m |\Phi\rangle \ , &
\bb K_m |\Phi\rangle & = K_m |\Phi\rangle \ , &
\bb H_m |\Phi\rangle & = H_m |\Phi\rangle \ , &
|\Phi\rangle & \in \{|\phi\rangle,\,|\psi\rangle\} \ ,
\end{align}
it is relatively easy to see that the relations
\begin{align}
& 2H_m = a_0 d_m + b_0 c_m = a_1 d_{m-1} + b_1 c_{m-1} \ , &&
2P_m = a_0 b_m + b_0 a_m = a_1 b_{m-1} + b_1 a_{m-1} \ , \nn
\\
&2 H_m = c_0 b_m + d_0 a_m = c_1 b_{m-1} + d_1 a_{m-1} \ , &&
2K_m = c_0 d_m + d_0 c_m = c_1 d_{m-1} + d_1 c_{m-1} \ ,
\end{align}
which follow from the graded commutation relations \eqref{eq:
suc(1|1)commrels}, can be used to solve recursively for the higher-level
representation parameters given their values at level 0 and 1.

As we will see, one property that does carry down from the higher-dimensional
cases is that the $\mathfrak{su}_c(1|1)$ R-matrix \eqref{redu}, \eqref{exacta},
\eqref{exacta1} exhibits an additional family of symmetries, $\mathbb{B}_n$, $n
\geq 1$, known as \emph{bonus} or \emph{secret}. These symmetries enhance the
$\mathcal Y(\su_c(1|1))$ Yangian to some \emph{indented} Yangian-like quantum
group we call $\mathcal Y(\gl_c(1|1))$, which contains all the generators in
\eqref{eq: suc(1|1)commrels} along with $\mathbb{B}_n$ (not including
$\mathbb{B}_0$).

\

To construct the RTT realisation of the Yangian we introduce a spectral
parameter $u$ defined in terms of $x^\pm$, and similarly $v$ for $y^\pm$. We
shall use two different definitions for the spectral parameter, but in both
cases the expansion of $x^\pm$ in powers of $u^{-1}$ will take the form
\begin{equation}\label{xpmu}
x^\pm = u + \mathcal{O}(1) \ .
\end{equation}
Once this expansion has been specified, we follow the construction outlined in
section \ref{sec2}. That is we expand the R-matrix $R(x^\pm,y^\pm)$ in inverse
powers of one of the two spectral parameters, say $u$,\footnote{For
definiteness we will assume that $M(x^+,x^-)= 2$. For the second set of
kinematical variables $y_\pm$, in terms of which we find the symmetry
generators, we will not assume anything and consequently the symmetries we find
hold for any $M$ in \eqref{eq:m}. If we leave $M(x^+,x^-)$ unfixed it simply
appears as an overall factor in the generators, for example $Q_m\,,\,S_m \sim
M(x^+,x^-)^{\frac12}$, and hence gives no new information.} and from the
resulting Laurent coefficients extract a series of generators $J_m$ whose
graded commutation relations reproduce those of the underlying
infinite-dimensional symmetry algebra. We can then define abstract generators
$\bb J_m$, of which $J_m$ are a representation. At this point we can construct
generators satisfying \eqref{eq: suc(1|1)commrels} along with their coproducts
and antipodes. It should be noted that this construction automatically leads to
the restricted Yangian, for which, in addition to \eqref{restrict}, the
higher-level central charges $\bb P_m$ and $\bb K_m$ are defined in terms of
lower-level central elements. We refer the reader to
\cite{Beisert:2014hya,Pittelli:2014ria} for a more complete discussion.

%%%%%%%%%%%%%%%%%%%%%%%%%%%%%%%%%%%%%%
\subsubsection{Canonical representation}\label{sec:can}
%%%%%%%%%%%%%%%%%%%%%%%%%%%%%%%%%%%%%%

Let us start by considering the canonical spectral parameter and a Hopf algebra
structure that is close in spirit to that used in the $AdS_5$ and $AdS_3$ cases
\cite{Beisert:2014hya,Pittelli:2014ria}. The spectral parameter is given by
\begin{equation}
u = \frac{1}{2}\big(x^+ + \frac{1}{x^+} + x^- + \frac{1}{x^-}\big) \ ,
\end{equation}
such that (assuming $M(x^+,x^-) = 2$) the expansions of $x^\pm$ are
\begin{equation}
x^\pm = u \pm \frac{i}{h} - \frac{1}{u} \pm \frac{i}{h u^2} + \mathcal{O}(u^{-3}) \ .
\end{equation}

Following the procedure outlined above we then identify the following
combinations\fn{Here the branch cut is chosen such that $\sqrt{i} =
e^{\f{i\pi}{4}}$ and $\sqrt{-i} = e^{-\f{i\pi}{4}}$.}
\ale{
&\bb Q_0\eqq \qfac\,\bb J^2_{01}\ , && \bb Q_1\eqq \qfac \,\bb J^2_{11}-\f i2\,\p{\id+\bb U^2}\sfac\,\bb J^1_{02}\ ,\nn\\
&\bb S_0\eqq \sfac\,\bb J^1_{02}\ , && \bb S_1\eqq \sfac\,\bb J^1_{12}+\f i2\p{\id+\bb U^{-2}}\qfac\,\bb J^2_{01}\ ,\nn\\
&\bb H_0\eqq \frac{ih}{2}\p{\bb J^1_{01}-\bb J^2_{02}}\ , && \bb H_1\eqq \f{ih}{2}\p{\bb J^1_{11}-\bb J^2_{12}} -\f {ih}4\p{\bb U^{2}-\bb U^{-2}}\ ,\nn\\
&\bb P_0\eqq \f h2\p{\id-\bb U^2} \ , && \bb P_1\eqq -i\p{\id +\bb U^2}\bb H_0\ , \nn \\
&\bb K_0\eqq \f h2\p{\id-\bb U^{-2}}\ , && \bb K_1\eqq i \p{\id +\bb U^{-2}}\bb H_0\ , \label{coprodcan}
}
which satisfy the defining commutation relations \eqref{eq: suc(1|1)commrels}.
Evaluated in the representation arising from the expansion of R-matrix, the
level-0 generators coincide with those used in \cite{Hoare:2014kma}.

\paragraph{Coproducts.}

The level-0 coproducts are given in \eqref{coprod0}, while the level-1
coproducts can be constructed from \eqref{eq: jsgeneralcoproducts} and read
\ale{
\Delta(\bb Q_1)&=\bb Q_1\otimes\id+\bb U\otimes\bb Q_1 \nn
\\\nn & \quad -\f ih\bb Q_0\otimes\bb H_0+\f ih\bb U\bb H_0\otimes\bb Q_0
+\f ih\bb U^2\bb S_0\otimes \bb P_0-\f ih\bb U^{-1}\bb P_0\otimes\bb S_0\ , \\
\nn \Delta(\bb S_1)&=\bb S_1\otimes\id+\bb U^{-1}\otimes\bb S_1
\\ & \quad +\f ih\bb S_0\otimes\bb H_0-\f ih\bb U^{-1}\bb H_0\otimes\bb S_0
- \f ih\bb U^{-2}\bb Q_0\otimes \bb K_0+\f ih\bb U\bb K_0\otimes\bb Q_0\ ,\nn
\\
\Delta(\bb H_1) & = \bb H_1 \otimes \id + \id \otimes \bb H_1 +
\f ih \bb U^{-2} \bb P_0 \otimes \bb K_0 + \f ih \bb U^2 \bb K_0 \otimes \bb P_0\ ,
}
where $\bb P_0$ and $\bb K_0$ are defined in terms of lower-level central
elements in \eqref{coprodcan}. Indeed inserting these definitions, the
coproduct for $\bb H_1$ becomes manifestly co-commutative as expected.

The coproducts for $\bb P_{1}$ and $\bb K_{1}$ can be obtained from the graded
commutation relations
\begin{equation}\label{coprodp1k1}
\Delta(\bb P_1) = \f12 \{\Delta(\bb Q_0),\Delta(\bb Q_1)\} \ ,\qquad
\Delta(\bb K_1) = \f12 \{\Delta(\bb S_0),\Delta(\bb S_1)\} \ .
\end{equation}
If $\bb P_1$ and $\bb K_1$ are defined in terms of lower-level central elements
as in \eqref{coprodcan} we find that these coproducts are also co-commutative
as required. {Moreover, we can compute} the coproducts for the level-2 central
charges
\begin{equation}\label{coprodp2k2}
\Delta(\bb P_2) = \f12 \{\Delta(\bb Q_1),\Delta(\bb Q_1)\} \ ,\quad
\Delta(\bb K_2) = \f12 \{\Delta(\bb S_1),\Delta(\bb S_1)\} \ , \quad
\Delta(\bb H_2) = \f12 \{\Delta(\bb Q_1),\Delta(\bb S_1)\} \ .
\end{equation}
Doing so, we find that for co-commutativity of $\Delta(\bb P_2)$ and
$\Delta(\bb K_2)$ we require
\ale{
& \bb P_2 \eqq -i\p{\id + \bb U^2}\bb H_1 - \f{1}{2h}\p{\id - \bb U^{2}} (\bb H_0^2 -\bb P_0 \bb K_0) \ ,\nn\\
& \bb K_2 \eqq i\p{\id + \bb U^{-2}} \bb H_1 - \f{1}{2h}\p{\id - \bb U^{-2}} (\bb H_0^2 - \bb P_0 \bb K_0) \ ,\label{co2c}}
where the normalisation is fixed by matching with the expansion of the
R-matrix,\footnote{It is worth recalling that the coproducts for the central
charges arising from the RTT realisation of the Yangian are co-commutative by
construction, and hence, up to a normalisation, $\bb P_2$ and $\bb K_2$ have to
take this form when evaluated in the representation arising from the expansion
of the R-matrix.} while the coproduct $\Delta(\bb H_2)$ is automatically
co-commutative upon using the definitions of $\bb P_{0,1}$ and $\bb K_{0,1}$ in
\eqref{coprodcan}.

With these definitions of the generators, one can show that the representation
of the Yangian arising from the expansion of the R-matrix is in general not
evaluation. However, if the eigenvalues of the central elements satisfy
\begin{equation}
H_0^2 - P_0 K_0 = 0 \ ,
\end{equation}
which has the interpretation as a (massless) shortening condition, we find that
the representation does become of evaluation type, in agreement with
\cite{Hoare:2014kma}. We will return to this issue in the following sections.

\paragraph{Crossing.}
Using the general expression for the antipodes of \eqref{antipodeexp} we can
derive the antipode for the generators of interest
\ale{
& \Sigma\comm{\bb Q_m}=-\bb U^{-1}\bb Q_m\ ,&&\Sigma\comm{\bb S_m}=-\bb U\bb S_m\ , \nn \\
& \Sigma\comm{\bb H_m}=-\bb H_m\ ,&& \Sigma\comm{\bb P_m}=\bb K_m\ ,\hspace{60pt} m=0,1 \ , \label{antipode}
}
i.e. it is involutive for the generators of $\mathcal Y_{0,1}(\su_c(1|1))$.

\paragraph{Secret symmetry.}

While $\bb J^1_{01}+\bb J^2_{02}$ is central, the combination
\be{
\mathbb{B}_1\eqq -\frac{ih}{2}\p{\bb J^1_{11}+\bb J^2_{12}} \ ,
}
satisfies
\ale{
\comm{\mathbb{B}_1,\bb Q_0}&=\bb Q_1 + i \p{\id+\bb U^2}\bb S_0\ , \qquad
\comm{\mathbb{B}_1,\bb S_0}=-\bb S_1 + i\p{\id+\bb U^{-2}}\bb Q_0\ ,
}
The coproduct reads
\be{
\Delta(\mathbb{B}_1)=\mathbb{B}_1 \otimes\id+\id\otimes\mathbb{B}_1-\f{i}{2h}\bb U^{-1}\bb Q_0\otimes\bb S_0-\f{i}{2h}\bb U\bb S_0\otimes \bb Q_0 \ ,
}
while the antipode for the secret symmetry is
\be{
\Sigma\comm{\mathbb{B}_1}=-\mathbb{B}_1 -\f i{h}\bb H_0\ .
}
As in the $AdS_5$ and $AdS_3$ case, the antipode is not an involution when
acting upon the secret symmetry.

%%%%%%%%%%%%%%%%%%%%%%%%%%%%%%%%%%%%%%
\subsubsection{Co-commutativity and shortening condition}\label{sec:cocom}
%%%%%%%%%%%%%%%%%%%%%%%%%%%%%%%%%%%%%%

Let us now investigate what happens if we try to impose evaluation
representation onto the Hopf algebra structure described in section
\eqref{sec:can}. In particular, we will show that this demonstrates the
existence of representations for which one of the higher-level central charges
is not co-commutative and hence does not admit an R-matrix.

In evaluation representation we have
\begin{eqnarray}
J_n = u^n J_0 \qquad \forall \, \, \, J \in \{Q,S,H,P,K\}\ ,
\end{eqnarray}
which manifestly satisfies the algebra relations \eqref{eq: suc(1|1)commrels}.
As discussed above, in order to have a co-commutative coproduct for all the
level-0 central charges one can impose
\begin{eqnarray}
P_0 = \frac{h}{2}(1-U^2)\ , \qquad K_0 = \frac{h}{2} (1 - U^{-2}) \ ,
\end{eqnarray}
where we recall that the constant $h$ is independent on the representation
space of the coproduct. Similarly, for the level-1 central charges for
co-commutativity one can impose
\begin{equation}
P_1 = -i (1+U^2)H_0 \ , \qquad K_1 = i (1+U^{-2}) H_0 \ ,
\end{equation}
It then follows that for a representation of evaluation type the spectral
parameter is given by
\begin{eqnarray}
u = -\frac{2i}{h} \frac{1+U^2}{1-U^2} H_0 \ ,
\end{eqnarray}
where we use $H_m$, $P_m$, $K_m$ and $U$ to denote both the generator in
evaluation representation and its eigenvalue, as it is always clear from
context which is meant.

At this point, we compute the coproduct of the level-2 central charge $P_2$
using
\begin{eqnarray}\label{coproddd}
\Delta(P_2) = \f12 \{\Delta(Q_1), \Delta(Q_1)\}\ .
\end{eqnarray}
The expression one obtains is rather lengthy, however it simplifies
considerably if one takes the antisymmetric combination
\begin{eqnarray}
\delta P_2 \equiv \Big(\Delta(P_2) - \Delta^{\text{op}} (P_2)\Big) \ ,
\end{eqnarray}
which is precisely the quantity that determines whether the coproduct is
co-commutative or not.

There are two notable contributions to $\delta P_2$, coming from two separate
pieces of the coproduct of $P_2$. The first contribution comes from the part of
$\Delta(P_2)$ arising when the $S_0$ generators in $\Delta(Q_1)$ meet among
themselves in the anti-commutator \eqref{coproddd}. Upon antisymmetrisation
these terms contribute
\begin{equation}
-\frac{1}{2h}\big((1-U^2) \otimes (1-U^2)\big) \big(P_0 K_0 \otimes 1 - 1 \otimes P_0 K_0\big) \ ,
\end{equation}
to $\delta P_2$. In fact this would be the only surviving term had we set
$H_0=u=0$ in both representation spaces. In principle this could already be
enough to conclude that there exist representations with non co-commutative
$P_2$. Nevertheless, it is instructive to continue.

The remaining terms reduce to
\begin{equation}
\frac{1}{2h}\big((1-U^2) \otimes (1-U^2)\big) \big(H_0^2 \otimes 1 - 1 \otimes H_0^2\big)\
\end{equation}
and, combining all contributions together, we obtain
\begin{eqnarray}
\delta P_2 =
\frac{1}{2h}\big((1-U^2) \otimes (1-U^2)\big) \big((H_0^2-P_0K_0) \otimes 1 - 1 \otimes (H_0^2-P_0K_0)\big)\ .
\end{eqnarray}
This means that we can achieve co-commutativity if we demand that
\begin{eqnarray}
\label{short}
H_0^2 - P_0K_0 = \mbox{constant} \ ,
\end{eqnarray}
where the constant does not depend on the representation space. The relation
\eqref{short} is nothing else than the known shortening condition, which is in
this way reinterpreted as the condition that makes the central charges'
coproduct co-commutative at higher levels (similarly to what the Serre
relations do for the canonical part of the Yangian).

%%%%%%%%%%%%%%%%%%%%%%%%%%%%%%%%%%%%%%
\subsubsection{Evaluation representation}\label{sec:eval}
%%%%%%%%%%%%%%%%%%%%%%%%%%%%%%%%%%%%%%

The Hopf algebra structure discussed in section \ref{sec:can}, which was
motivated by similar constructions in the $AdS_5$ and $AdS_3$ cases, turned out
to give a representation of the Yangian that was not of evaluation type. It
turns out that there is an alternative Hopf algebra structure we can put on the
same infinite-dimensional algebra, such that the representation arising from
the expansion of the R-matrix is evaluation.

To do this we introduce a new spectral parameter
\ale{\label{eq: specparevrep}
&u\eqq \frac14 \big(1+ \sqrt{\frac{x^+}{x^-}}\big)^2 \big(x^- + \frac{1}{x^+}\big)
= \frac14 \big(1+ \sqrt{\frac{x^-}{x^+}}\big)^2 \big(x^+ + \frac{1}{x^-}\big) \ ,
}
such that the expansions of $x^\pm$ (again assuming that $M(x^+,x^-) = 2$) are
given by
\be{\label{eq: xplusminusofu}
x^\pm\eqq u\pm \f ih-\big(1+\f 1{4h^2}\big)\f1u\pm \f i{h\, u^2}+\mathcal{O}(u^{-3}) \ ,
}

Let us now define the following (as an alternative to \eqref{coprodcan})
combinations of generators
\ale{
&\bb Q_0\eqq \qfac\,\bb J^2_{01}\ , && \bb Q_1\eqq \qfac \,\bb J^2_{11}- i\bb U\sfac\,\bb J^1_{02}\ ,\nn\\
&\bb S_0\eqq \sfac\,\bb J^1_{02}\ , && \bb S_1\eqq \sfac\,\bb J^1_{12}+i\bb U^{-1}\qfac\,\bb J^2_{01}\ ,\nn\\
&\bb H_0\eqq \frac{ih}{2}\p{\bb J^1_{01}-\bb J^2_{02}}\ , && \bb H_1\eqq \f{ih}{2}\p{\bb J^1_{11}-\bb J^2_{12}} -\f {ih}2\p{\bb U-\bb U^{-1}}\ ,\nn\\
&\bb P_0\eqq \f h2\p{\id-\bb U^2} \ , && \bb P_1\eqq -\frac i2 \p{\id +\bb U}^2\bb H_0\ , \nn \\
&\bb K_0\eqq \f h2\p{\id-\bb U^{-2}}\ , && \bb K_1\eqq \f i2 \p{\id +\bb U^{-1}}^{2}\bb H_0\ , \label{coprodeval}
}
which also satisfy the defining commutation relations \eqref{eq:
suc(1|1)commrels}. Again, evaluated in the representation arising from the
expansion of R-matrix, the level-0 generators coincide with those used in
\cite{Hoare:2014kma}.

\paragraph{Coproducts.}

The level-0 coproducts are given in \eqref{coprod0}, while the level-1
coproducts can be constructed from \eqref{eq: jsgeneralcoproducts} and read
\ale{
\Delta(\bb Q_1)&=\bb Q_1\otimes\id+\bb U\otimes\bb Q_1 \nn
\\\nn & \quad -\f ih\bb Q_0\otimes\bb H_0+\f ih\bb U\bb H_0\otimes\bb Q_0
+i\bb U\bb S_0\otimes (\id - \bb U) - i (\id - \bb U)\otimes\bb U \bb S_0\ , \\
\nn \Delta(\bb S_1)&=\bb S_1\otimes\id+\bb U^{-1}\otimes\bb S_1
\\ & \quad +\f ih\bb S_0\otimes\bb H_0-\f ih\bb U^{-1}\bb H_0\otimes\bb S_0
-i\bb U^{-1}\bb Q_0\otimes (\id - \bb U^{-1}) + i (\id - \bb U^{-1})\otimes\bb U^{-1} \bb Q_0\ , \nn \\
\Delta(\bb H_1) & = \bb H_1 \otimes \id + \id \otimes \bb H_1 \nn
\\ & \quad -
\f {ih}2\big(\bb U \otimes \bb U - \bb U^{-1} \otimes \bb U^{-1} - (\bb U - \bb U^{-1})\otimes \id
-\id \otimes (\bb U - \bb U^{-1}) \big) \ .
}
Here $\Delta(\bb H_1)$ is written in a manifestly co-commutative form. To check
that these coproducts, along with those in \eqref{coprod0}, obey the graded
commutation relations \eqref{eq: suc(1|1)commrels} one needs to use the
definitions of $\bb P_0$ and $\bb K_0$ in \eqref{coprodeval}.

The coproducts for $\bb P_{1}$ and $\bb K_{1}$ can be obtained from the graded
commutation relations \eqref{coprodp1k1}. If $\bb P_1$ and $\bb K_1$ are
defined in terms of lower-level central elements as in \eqref{coprodeval} we
find that these coproducts are also co-commutative as required. Furthermore, we
can compute the coproducts for the level-2 central charges using
\eqref{coprodp2k2}. Co-commutativity of $\Delta(\bb P_2)$ and $\Delta(\bb
K_2)$ then requires
\ale{
& \bb P_2 \eqq -2i \bb U \bb H_1 - \f{1}{2h}\p{\id - \bb U^{2}} \bb H_0^2 \ ,\nn\\
& \bb K_2 \eqq 2i \bb U^{-1} \bb H_1 - \f{1}{2h}\p{\id - \bb U^{-2}} \bb H_0^2 \ ,\label{co2e}}
where normalisations are fixed by matching with the expansion of the R-matrix,
while the coproduct $\Delta(\bb H_2)$ is automatically co-commutative by the
definitions of $\bb P_{0,1}$ and $\bb K_{0,1}$ in \eqref{coprodeval}.

From \eqref{coprodeval} we find that the representation of the Yangian arising
from the R-matrix expansion is indeed of evaluation type with spectral
parameter \eqref{eq: specparevrep}
\begin{equation}
u = -\frac{i}{h}\frac{1+U}{1-U}H_0 \ .
\end{equation}

\paragraph{Crossing.}
Using the general expression for the antipodes of \eqref{antipodeexp} we can
derive the antipode for the generators of interest. These turn out to be the
same as for the canonical case, i.e. \eqref{antipode}.

\paragraph{Secret symmetry.}

While $\bb J^1_{01}+\bb J^2_{02}$ is central, the combination
\be{
\mathbb{B}_1\eqq -\frac{ih}{2}\p{\bb J^1_{11}+\bb J^2_{12}} \ ,
}
satisfies
\ale{
\comm{\mathbb{B}_1,\bb Q_0}&=\bb Q_1 + \frac i2 \p{\id+\bb U}^2\,\bb S_0 \ ,\qquad
\comm{\mathbb{B}_1,\bb S_0}=-\bb S_1 +\f i2\p{\id+\bb U^{-1}}^2\,\bb Q_0\ .
}
The coproduct reads
\be{
\Delta(\mathbb{B}_1)=\mathbb{B}_1 \otimes\id+\id\otimes\mathbb{B}_1-\f{i}{2h}\bb U^{-1}\bb Q_0\otimes\bb S_0-\f{i}{2h}\bb U\bb S_0\otimes \bb Q_0 \ ,
}
while the antipode for the secret symmetry is
\be{
\Sigma\comm{\mathbb{B}_1}=-\mathbb{B}_1 -\f i{h}\bb H_0\ .
}
As before, the antipode is not an involution when acting upon the secret symmetry.

It is worth highlighting that the modification of the combinations in
\eqref{coprodeval} has altered the commutation relations involving $\bb B_1$,
changing the tail.

%%%%%%%%%%%%%%%%%%%%%%%%%%%%%%%%%%%%%%
\subsubsection{Freedom in the realisation of the Yangian}\label{sec:freedom}
%%%%%%%%%%%%%%%%%%%%%%%%%%%%%%%%%%%%%%

The two different Hopf algebra structures described in sections \ref{sec:can}
and \ref{sec:eval} are indicative of a larger possible freedom, which we will
now describe. Let us consider the defining graded commutation relations
\eqref{eq: suc(1|1)commrels}, but in particular focus on the restricted form in
which $\bb P_m$ and $\bb K_m$ are defined in terms of lower-level central
charges.

Motivated by the definitions of $\bb P_{0,1,2}$ and $\bb K_{0,1,2}$ in
\eqref{coprodcan}, \eqref{co2c}, \eqref{coprodeval}, \eqref{co2e}, we postulate
that the following relation is true for all levels:
\begin{equation}\label{constraints}
\bb U^{-1} \bb P_m = - \bb U \bb K_m\ .
\end{equation}
\def\Xy{{\bb Y}}
\def\Xz{{\bb Z}}
If we now consider the following redefinitions:
\begin{align}
\tilde{\bb Q}_0 & = \bb Q_0\ , \qquad \
\tilde{\bb S}_0 = \bb S_0\ , \qquad \ \,
\tilde{\bb P}_0 = \bb P_0\ , \qquad \
\tilde{\bb K}_0 = \bb K_0\ , \qquad \
\tilde{\bb H}_0 = \bb H_0\ ,\nn
\\ \tilde{\bb Q}_m & = \bb Q_m + \sum_{k=0}^{m-1} y_{q_{m,k}} \bb Q_{m-k} + z_{q_{m,k}} \bb S_{m-k}\ , \qquad
\tilde{\bb S}_m = \bb S_m + \sum_{k=0}^{m-1} y_{s_{m,k}} \bb S_{m-k} + z_{s_{m,k}} \bb Q_{m-k}\ ,\nn
\\ \tilde{\bb P}_m & = \bb P_m + \delta^{\bb P}_m\ , \qquad\quad\!
\tilde{\bb K}_m = \bb K_m + \delta^{\bb K}_m\ , \qquad\quad\!
\tilde{\bb H}_m = \bb H_m + \delta^{\bb H}_m\ , \qquad\quad\! m > 1\ ,\label{gentrans}
\end{align}
where $y_{q,s}$ and $z_{q,s}$ are functions of the braiding factor $\bb U$. We
are interested in finding a set of these functions such that the algebra
relations \eqref{eq: suc(1|1)commrels} and \eqref{constraints} are still
satisfied by the new generators. Indeed such a solution exists and is given by
\begin{align}
y_{_{m,k}} &= y_{q_{m,k}} = y_{s_{m,k}} =
\frac12 \binom{m}{k} \left(\left(y + z\right)^{m-k} + \left( y - z\right)^{m-k} \right)\ , \nn \\
z_{_{m,k}} & =-i \bb U^{-1} z_{q_{m,k}} = i \bb U z_{s_{m,k}} =
\frac12 \binom{m}{k} \left(\left(y + z\right)^{m-k} - \left( y - z\right)^{m-k} \right)\ , \nn
\\ \delta^{\bb P}_m & = \sum_{k=0}^{m-1} y_{_{m,k}} \bb P_k + i \bb U z_{_{m,k}} \bb H_k\ , \qquad
\delta^{\bb K}_m = \sum_{k=0}^{m-1} y_{_{m,k}} \bb K_k - i \bb U^{-1} z_{_{m,k}} \bb H_k\ , \nn
\\ \delta^{\bb H}_m & = \sum_{k=0}^{m-1} y_{_{m,k}} \bb H_k + i \bb U z_{_{m,k}} \bb K_k
= \sum_{k=0}^{m-1} y_{_{m,k}} \bb H_k - i \bb U^{-1} z_{_{m,k}} \bb P_k\ .\label{gensol}
\end{align}
I.e. the freedom is parameterised by two functions, $y$ and $z$, of the
braiding factor $\bb U$. The freedom parameterised by $y$ is a generalisation
of the symmetry \eqref{eq: yangianshift}. For a representation of evaluation
type, its effect is to shift the spectral parameter by $y(U)$.

The redefinitions \eqref{gentrans} will modify many of the relations underlying
the Hopf algebra structure, including the relations between $\bb P_m$ and $\bb
K_m$ and the lower-level central charges, the coproducts of the generators and
the commutation relations involving the secret symmetry $\bb B_1$. If we
demand that the antipode structure \eqref{antipode} is preserved, we find
\begin{equation}
\Sigma(y(\bb U)) = y(\bb U) \ , \qquad \Sigma(z(\bb U)) = z(\bb U) \ .
\end{equation}
These relations are solved by functions symmetric in $\bb U$ and $\bb U^{-1}$.

Observing that mapping between the Hopf algebra structures in sections
\ref{sec:can} and \ref{sec:eval} precisely takes the form given above in
\eqref{gentrans} and \eqref{gensol}, we investigate what happens if we take a
more general ansatz for the level-1 Yangian supercharges
\ale{
\bb Q_0&\eqq \qfac \, \bb J^2_{01} \ ,
&\bb S_0&\eqq \sfac \, \bb J^1_{02} \ ,\nn
\\\label{eq: oneparfamilyofsuperlevelonecharges}
\bb Q_1&\eqq \qfac \, \bb J^2_{11} + i \bb U z(\bb U) \, \sfac \bb J^1_{02} \ ,
&\bb S_1&\eqq \sfac \, \bb J^1_{12} - i \bb U^{-1} z(\bb U) \, \qfac \bb J^2_{01} \ ,
}
with
\begin{equation}
z(\bb U) = z(\bb U^{-1}) \ ,
\end{equation}
to preserve the antipode structure. By anti-commuting $\bb Q_1$ and $\bb S_1$
we obtain the central charges
\ale{\label{eq: generalcc}
\bb P_1 & = -\f i2 \p{\id -2 \bb U z(\bb U) + \bb U^2}\bb H_0 \ , \nn && \hspace{-20pt}
\bb P_2=2i\bb U z(\bb U)\,\bb H_1 - h^{-2} \p{\bb H_0^2 - h^2(1- z (\bb U)^2)}\bb P_0 \ , \nn
\\ \bb K_1 & = \f i2 \p{\id -2 \bb U^{-1} z (\bb U) + \bb U^{-2}} \bb H_0 \ , && \hspace{-20pt}
\bb K_2=-2i\bb U^{-1} z(\bb U)\,\bb H_1 - h^{-2} \p{\bb H_0^2 - h^2(1- z (\bb U)^2)}\bb K_0 \ , \nn
\\ \bb H_1 & = \f{ih}{2}\p{\bb J^{1}_{11} - \bb J^{2}_{12}} + \f{ih}{2}\p{\bb U - \bb U^{-1}}z(\bb U) \ . &&
}

We can now ask for what choices of the function $z(\bb U)$ we can have an
evaluation type representation of the Yangian. In particular, this would imply
the following two relations
\be{\label{eq: cocommutativityandevrepforp2}
P_0\, P_2= P_1^2 \ , \qquad
P_0\, H_1= P_1\, H_0\ .
}
Combining \eqref{eq: generalcc} with the conditions just above reveals that a
necessary requirement for an evaluation type representation is
\be{\label{eq: disprelfromyangian}
U^2 (1-z(U)^2)(H_0^2 - P_0 K_0) = 0 \ .
}

There are two cases solutions of interest to this condition. The first is
\be{\label{disprelfromyangian}
H_0^2 - P_0K_0 =0 \ ,
}
which can be interpreted as a (massless) shortening conditon. This is indeed
consistent with our findings in section \ref{sec:can}. If we admit long
representations, as in the context of the $\mathfrak{su}_c(1|1)$ R-matrix
\eqref{redu}, \eqref{exacta}, \eqref{exacta1}, a consistent evaluation
representation demands
\begin{equation} z(\bb U)= \pm \id \ .
\end{equation}
Indeed, the choice $z(\bb U)\eqq -\id$ was the representation analysed in
section \ref{sec:eval}.

Finally, let us observe that the generator
\begin{equation}
\bb B_1 = -\frac{ih}{2} \p{\bb J^1_{11} + \bb J^2_{12}} \ ,
\end{equation}
now satisfies
\ale{
\comm{\mathbb{B}_1,\bb Q_0}&=\bb Q_1 + \frac i2 \p{\id-2\bb U z(\bb U) + \bb U^2}\,\bb S_0 \ ,\nn \\
\comm{\mathbb{B}_1,\bb S_0} &=-\bb S_1 +\f i2\p{\id-2\bb U^{-1} z(\bb U) + \bb U^{-2}}\,\bb Q_0\ . \label{bcomgen}
}
Choosing
\begin{equation}\label{zch}
z(\bb U) = \frac{1}{2} \p{\bb U + \bb U^{-1}} \ ,
\end{equation}
we see that $\bb P_1$ and $\bb K_1$ in \eqref{eq: generalcc}, along with the
tails in \eqref{bcomgen} vanish. It is therefore natural to ask if the
existence of this choice \eqref{zch} is related to the existence of the secret
symmetry.

%%%%%%%%%%%%%%%%%%%%%%%%%%%%%%%%%%%%%%%%
\section{Strong and weak coupling expansions}\label{sec4}
%%%%%%%%%%%%%%%%%%%%%%%%%%%%%%%%%%%%%%%%

%%%%%%%%%%%%%%%%%%%%%%%%%%%%%%%%%%%%%%
\subsection{Strong coupling expansion and the classical $r$-matrix}
%%%%%%%%%%%%%%%%%%%%%%%%%%%%%%%%%%%%%%

As was done in the $AdS_5$ case \cite{Torrielli:2007mc,Beisert:2007ty} it is
instructive to study the so-called {\it classical $r$-matrix} of the system.
This can be obtained by expanding the quantum R-matrix
\be{
R=\id\otimes\id+h^{-1}r+O\p{h^{-2}} \ ,
}
at strong coupling. In standard quantum group theory, the knowledge of the
classical $r$-matrix and of its Lie bi-algebra structure allows one to
reconstruct the quantum group underlying the exact problem. This is still an
open problem for $AdS$ superstrings, nevertheless much can be learnt from this
exercise.

%%%%%%%%%%%%%%%%%%%%%%%%%%%%%%%%%%%%%%
\subsubsection{Parameterisation and loop algebra}
%%%%%%%%%%%%%%%%%%%%%%%%%%%%%%%%%%%%%%

Following \cite{Arutyunov:2006iu} we introduce $\zeta = h^{-1}$ and the
spectral paramter $z$
\be{
x^\pm\eqq z\p{\sqrt{1-\f{\zeta^2}{\p{z-\f1z}^2}}\pm\f{i\zeta }{z-\f1z}} \ ,
}
where as before we assume $M(x^+,x^-) = 2$. Expanding the representations of
the generators we find
\ale{
U&=\exp\p{i\zeta \mathfrak D}=\id + i\zeta \mathfrak D+\mathcal{O}(\zeta^2) \ ,\nn\\
P_0&=-i\mathfrak D+\mathcal{O}(\zeta)\ ,\qquad K_0=i\mathfrak D+\mathcal{O}(\zeta),\qquad \ \ \, H_0=\mathfrak H_0+\mathcal{O}(\zeta)\ ,\nn\\
Q_0&=\mathfrak Q_0+\mathcal{O}(\zeta)\ ,\qquad \ \ \, S_0=\mathfrak S_0+\mathcal{O}(\zeta) \ ,\nn\\
Q_1&=\mathfrak \zeta^{-1}Q_1+\mathcal{O}(\zeta)\ ,\quad \ S_1=\zeta^{-1} \mathfrak S_1+\mathcal{O}(\zeta) \ ,\quad
B_1=\zeta^{-1}\mathfrak B_1+\mathcal{O}(1)\ .
}

The $\zeta\to0$ limit of the spectral parameter \eqref{eq: specparevrep} is
\be{
u\eqq \frac14 \big(1+ \sqrt{\frac{x^+}{x^-}}\big)^2 \big(x^- + \frac{1}{x^+}\big)
= \frac14 \big(1+ \sqrt{\frac{x^-}{x^+}}\big)^2 \big(x^+ + \frac{1}{x^-}\big)
\to z+z^{-1} \ .
}
In what follows, it is convenient to perform the rescaling
\be{
u\to \frac{u}{2i} \ ,
}
such that the limiting generators $\mathfrak J$ are in the evaluation
representation with
\be{
\mathfrak J_m=u^m\mathfrak J_0 \ ,\qquad u=\f1{2i}(z+z^{-1})\equiv -i\mathfrak H_0\mathfrak D^{-1} \ .
}
The non-trivial commutation relations for these $\mathfrak J_m$ read
\ale{
& \acomm{\mathfrak Q_m,\mathfrak Q_n}=-\acomm{\mathfrak S_m,\mathfrak S_n}=2\mathfrak H_{m+n-1} \ , && \acomm{\mathfrak Q_m,\mathfrak S_n}=2\mathfrak H_{m+n}\ ,\nn\\
& \comm{\mathfrak B_m,\mathfrak Q_n}=\mathfrak Q_{m+n}+\mathfrak S_{m+n-1}, &&\comm{\mathfrak B_m,\mathfrak S_n}=-\mathfrak S_{m+n}+\mathfrak Q_{m+n-1} \ .
}

%%%%%%%%%%%%%%%%%%%%%%%%%%%%%%%%%%%%%%
\subsubsection{Classical $r$-matrix for the deformed $\gl(1|1)_{u,u^{-1}}$}
%%%%%%%%%%%%%%%%%%%%%%%%%%%%%%%%%%%%%%

The classical limit of the R-matrix gives the classical $r$-matrix, whose
non-trivial entries are\footnote{In this section we label the spectral
parameters with integers, for example $z_1$, $z_2$, $z_3$ and so on.}
\ale{
& r_{14}=r_{41}=-\frac{i \sqrt{\frac{{z_1}^2}{{z_1}^2-1}} \sqrt{\frac{{z_2}^2}{{z_2}^2-1}}}{{z_1} {z_2}-1}\ , &&r_{23}=r_{32}=\f{{z_1} {z_2}-1}{z_1-z_2}\,r_{14}\ ,\nn\\
& r_{22}=\frac{i {z_1}^2 \left({z_2}^2-1\right)}{\left({z_1}^2-1\right) ({z_1}-{z_2}) ({z_1} {z_2}-1)}\ , && r_{33}=\frac{i \left({z_1}^2-1\right) {z_2}^2}{\left({z_2}^2-1\right) ({z_1}-{z_2}) ({z_1} {z_2}-1)}\ ,\nn\\
& r_{44}=\frac{i \left({z_1}^2 \left(\left({z_1}^2-4\right) {z_2}^2+{z_2}^4+1\right)+{z_2}^2\right)}{\left({z_1}^2-1\right) \left({z_2}^2-1\right) ({z_1}-{z_2}) ({z_1} {z_2}-1)}\ . &&
}
The residue at $z_2= z_1$ is
\be{
\text{Re}\,r_{|_{z_2\to z_1}}=f(z_1)\p{\id\otimes\id-\mathfrak C}=\f{i z_1^2}{1-z_1^2}\p{\begin{matrix}
0 & 0 & 0 & 0 \\
0 & 1 & -1 & 0 \\
0 & -1 & 1 & 0 \\
0 & 0 & 0 & 2
\end{matrix}},
}
where $\mathfrak C$ is the Casimir operator for the $\gl(1|1)$ tensor algebra.

The classical $r$-matrix admits the following expression in terms of the
generators of the $\gl(1|1)$ algebra
\ale{
r&=\f1{u_2-u_1}\Big(\f12\mathfrak Q_0\otimes \mathfrak S_0-\f12\mathfrak S_0\otimes\mathfrak Q_0-\f{u_1}{u_2}\mathfrak B_0\otimes \mathfrak H_0 -\f{u_2}{u_1}\mathfrak H_0\otimes \mathfrak B_0\nn
\\ & \hspace{200pt} +\f{1+u_1^2+u_2^2}{u_1 u_2}\,\mathfrak H_0\otimes \mathfrak H_0\Big) \ .
}
The same matrix can then be rewritten as an element in the tensor product of
two copies of the loop algebra $\gl(1|1)_{u,u^{-1}}$
\ale{\label{eq: classicalrmatrixrepindependent}
r&=r_{\psu(1|1)}-\sum_{n=0}^\infty\p{\tilde{\mathfrak B}_{n+1}\otimes \mathfrak H_{-n-2}+{\mathfrak H}_{n-1}\otimes \tilde{\mathfrak B}_{-n}-\mathfrak H_{n-1}\otimes \mathfrak H_{-n-2}} \ ,
}
where
\be{
r_{\psu(1|1)}\eqq \f12\sum_{n=0}^\infty\p{\mathfrak Q_n\otimes \mathfrak S_{-n-1}-\mathfrak S_n\otimes\mathfrak Q_{-n-1}} \ ,\qquad \tilde{\mathfrak B}_{n}\eqq {\mathfrak B}_{n}-{\mathfrak H}_{n}\ .
}
The peculiarity of \eqref{eq: classicalrmatrixrepindependent} is that it is
representation independent, and can therefore be taken as a candidate for the
{\it universal} classical $r$-matrix in the $AdS_2$ case.

%%%%%%%%%%%%%%%%%%%%%%%%%%%%%%%%%%%%%%
\subsection{Weak coupling limit and Bethe equations}
%%%%%%%%%%%%%%%%%%%%%%%%%%%%%%%%%%%%%%

In this section we study the leading-order weak-coupling ($h \to 0$) term in
the R-matrix, extracting from it a set of Bethe equations. These equations
should relate to the leading-order first-level nested Bethe equations one
would in principle obtain from the spin-chain Hamiltonian of the putative dual
superconformal quantum mechanics that is meant to live on the boundary of
$AdS_2$. In $AdS_5$ parlance, this would be called the {\it one-loop}
nearest-neighbour spin-chain Hamiltonian \cite{Minahan:2002ve}.

The advantage of restricting to this limit is that we can avoid one crucial
complication present when dealing with the full R-matrix. To admit a
pseudo-vacuum we need to take the tensor product of two copies of the
centrally-extended $\mathfrak{su}(1|1)$ R-matrix. (This tensor product is the
one relevant for building up the $AdS_2 \times S^2$ worldsheet S-matrix
\cite{Hoare:2014kma}). The corresponding pseudo-vacuum is a specific fermionic
linear combination of the states in the two copies, with a definite charge
under a certain $U(1)$ quantum number. The corresponding $U(1)$ symmetry does
not act in a well-defined way on the individual copies. Performing the
algebraic Bethe-ansatz procedure starting from the full pseudo-vacuum is at the
moment an open issue.

Dealing with the individual copies, which do not admit a pseudo-vacuum in their
own right, could in principle be approached by adapting alternative methods
(such as, for instance, the one of Baxter operators). However, the limit $h \to
0$ switches off the most unconventional entries of the R-matrix and allows for
the existence of a pseudo-vacuum separately in each copy. Moreover, it
drastically simplifies all the remaining entries, allowing for an almost
straightforward treatment.

%%%%%%%%%%%%%%%%%%%%%%%%%%%%%%%%%%%%%%
\subsubsection{Weak coupling R-matrix}
%%%%%%%%%%%%%%%%%%%%%%%%%%%%%%%%%%%%%%
The R-matrix up to order $h$ has the form
\be{\label{as}
R_{12}=R_{12}^{(0)}+h \,R_{12}^{(1)} \ ,
}
where
\be{
R_{12}^{(0)}=\p{\begin{matrix}
1 & 0 & 0 & 0 \\
0 & B_{12} & C_{12} & 0 \\
0 & C_{12} & D_{12} & 0 \\
0 & 0 & 0 & E_{12}
\end{matrix}}\ ,\qquad \qquad R_{12}^{(1)}=\p{\begin{matrix}
0 & 0 & 0 & A_{12} \\
0 & 0 & 0 & 0 \\
0 & 0 & 0 & 0 \\
A_{12} & 0 & 0 & 0
\end{matrix}}\ ,\label{matr}
}
with the parameterising functions given by
\ale{
A_{12}&\eqq \f{4i(u_1-u_2)(u_1u_2-1)}{(1+u_1^2)(1+u_2^2)(u_1-u_2-2i)}e^{-\f{i}4\p{p_1-p_2}} \ ,\nn\\
B_{12}&\eqq \f{u_1-u_2}{u_1-u_2-2i}e^{-\f{i}2p_1}\ ,\nn\\
C_{12}&\eqq \f{2i}{u_1-u_2-2i}e^{-\f{i}4(p_1-p_2)}\ ,\nn \\
D_{12}&\eqq \f{u_1-u_2}{u_1-u_2-2i}e^{\f{i}2p_2}\ ,\nn\\
E_{12}&\eqq \f{u_1-u_2+2i}{u_1-u_2-2i}e^{-\f{i}2(p_1-p_2)} \ .\label{pff}
}
The maps between $x^\pm$, $u$ and $p$ are
\be{
x^\pm = \f{1}{2h}\Big(\f{\text{cn}(\f p2,-4h^2)}{\text{sn}(\f p2,-4h^2)}\pm i\Big)\Big(1+\text{dn}(\f p2,-4h^2)\Big) \ ,
\qquad u = \cot \f p4 \ .
}

The factors of $e^{ip_k}$ in \eqref{matr} and \eqref{pff} are the result of a
Drinfeld twist \cite{Drinfeld:1989st}. The presence of such a twist was
already observed in the same scaling limit of the $AdS_5$ R-matrix \cite{rev}.
Indeed, $R_{12}$ can be written as
\be{
R_{12}=T_{21}\tilde R_{12} T^{-1}_{12}\ ,
}
where
\be{
T_{12}\eqq \text{diag}
\big(e^{\alpha(p_1+p_2)} , \,
e^{-\f{i}4p_2+\beta(p_1+p_2)}, \,
e^{(\beta-\f{i}4)p_1+(\beta-\f{i}2)p_2}, \,
e^{(\alpha+\f{i}4)p_1+(\alpha-\f{i}4)p_2}\big) \ ,\nn
}
with $\alpha$ and $\beta$ arbitrary coefficients. The entries of $\tilde
R_{12}$ are given by
\ale{
\tilde A_{12}&\eqq \f{4i(u_1-u_2)(u_1u_2-1)}{(1+u_1^2)(1+u_2^2)(u_1-u_2-2i)} \ ,\nn\\
\tilde B_{12}&\eqq \f{u_1-u_2}{u_1-u_2-2i} \ ,\nn\\
\tilde C_{12}&\eqq \f{2i}{u_1-u_2-2i}\ , \nn\\
\tilde D_{12}&\eqq \f{u_1-u_2}{u_1-u_2-2i}\ ,\nn\\
\tilde E_{12}&\eqq \f{u_1-u_2+2i}{u_1-u_2-2i} \ ,
}
with the same associations of letters to entries as in \eqref{as} and
\eqref{matr}.

Taking $h = 0$, the entries of $\tilde R_{12}$ involving $\tilde A_{12}$ drop
out, and we recover the $\mathfrak{gl}(1|1)$ R-matrix, written down in its
canonical rational form.

%%%%%%%%%%%%%%%%%%%%%%%%%%%%%%%%%%%%%%
\subsubsection{Bethe ansatz and twist}
%%%%%%%%%%%%%%%%%%%%%%%%%%%%%%%%%%%%%%

As we have seen, at leading order the R-matrix reduces to the canonical
rational (Yangian) R-matrix $R_{\text{can}}$ of $\mathfrak{gl}(1|1)$, decorated
by a Drinfeld twist. Let us rewrite the twist as
\begin{eqnarray}
\label{twist}
T_{12} = t_{12}^{ij} \, E_{ii} \otimes E_{jj} \, = \, e^{i[\underline{i} \, p_1 + \underline{j} \, p_2]} \, E_{ii} \otimes E_{jj} \ ,
\end{eqnarray}
where $E_{ij}$ are unit matrices, i.e. $1$ in row $i$, column $j$ and zero
everywhere else, and we denote by $\underline{i}$ the numerical coefficient
multiplying the momentum in the respective spaces of the $T_{12}$ matrix.
Similarly, we write the canonical R-matrix as
\begin{eqnarray}
\label{Rmat}
R_{\text{can}}{}_{12} = r_{ij}^{kl}(p_1,p_2) \, E_{ki} \otimes E_{jl} \, \equiv \, r_{12}{}_{ij}^{kl}\, E_{ki} \otimes E_{jl} \ ,
\end{eqnarray}
such that at leading order the R-matrix reads
\begin{eqnarray}
R_{12} = T_{21} \, R_{\text{can}}{}_{12} \, T_{12}^{-1} \ .
\end{eqnarray}

We are now ready to write down the monodromy matrix. Denoting the auxiliary
space with the label $0$, we have
\begin{align}
M & = R_{01} \, R_{02} \, \cdots \, R_{0N} \nn \\
& = T_{10} \, R_{\text{can}}{}_{01} \, T_{01}^{-1} \, T_{20}
\, R_{\text{can}}{}_{02} \, T_{02}^{-1} \,
\cdots \, T_{N0} \, R_{\text{can}}{}_{0N} \, T_{0N}^{-1}\ .\label{inin}
\end{align}
Plugging the explicit expressions \eqref{twist} and \eqref{Rmat} into
\eqref{inin} we obtain
\begin{align}
M & =t_{10}^{i_1 j_1} \, t_{20}^{i_3 k_1} \, t_{30}^{i_5 k_2} \, \cdots \, r_{01}{}_{k_1 j_2}^{j_1 i_1} \,
r_{02}{}_{k_2 j_4}^{k_1 i_3} \, r_{03}{}_{k_3 j_6}^{k_2 i_5} \, \cdots \,
(t_{01}^{k_1 j_2})^{-1}(t_{02}^{k_2 j_4})^{-1}(t_{03}^{k_3 j_6})^{-1}\, \cdots \nn
\\ & \hspace{220pt} E_{j_1 k_N} \otimes E_{i_1 j_2} \otimes E_{i_3 j_4} \otimes E_{i_5 j_6} \cdots \ .
\end{align}
Now we insert the dependence of the twist on the momentum. Most of the factors
appearing in the auxiliary space cancel, such that we are left with
\begin{align}
M & = e^{i [\underline{j_1} - \underline{k_N}] p_0} \, e^{i [\underline{i_1} - \underline{j_2}] p_1} \,e^{i [\underline{i_3} - \underline{j_4}] p_2} \,e^{i [\underline{i_5} - \underline{j_6}] p_3} \cdots
r_{01}{}_{k_1 j_2}^{j_1 i_1} \, r_{02}{}_{k_2 j_4}^{k_1 i_3} \, r_{03}{}_{k_3 j_6}^{k_2 i_5} \, \cdots
\\& \hspace{200pt} E_{j_1 k_N} \otimes E_{i_1 j_2} \otimes E_{i_3 j_4} \otimes E_{i_5 j_6} \cdots \ . \nonumber
\end{align}
We can therefore define a new set of states and matrices
\begin{eqnarray}
\ses_{ab} \equiv e^{i [\underline{a} - \underline{b}] p} \, E_{ab} \ ,
\qquad |\underline{a}\rangle \equiv e^{i \underline{a} p} \, |a\rangle\ ,
\end{eqnarray}
such that we still have
\begin{eqnarray}
\ses_{ab} \, \ses_{cd} \, = \, \delta_{bc} \, \ses_{ad} \ , \qquad
\ses_{ab} \, |\underline{c}\rangle = \delta_{bc} \, |\underline{a}\rangle \ , \qquad
\langle\underline{a}|\underline{b}\rangle = \delta_{ab}\ .
\end{eqnarray}
Using these new vectors and unit matrices, the expression for $M$ becomes
indistinguishable from the canonical one. Consequently the Bethe ansatz
reduces to the standard one, except with the vectors $|\underline{a}\rangle$
now appearing in the wave functions: in particular, the Bethe equations for $M$
magnons will be
\be{\label{eq: beforffmagn}
\Big(\f{u_k-i}{u_k+i}\Big)^L=(-1)^{M-1} \ .
}
In the effective model obtained from this limit, excitations propagate as free fermions on a periodic one-dimensional
lattice.

%%%%%%%%%%%%%%%%%%%%%%%%%%%%%%%%%%%%%%
\subsubsection{Bethe equations via the algebraic Bethe ansatz}
%%%%%%%%%%%%%%%%%%%%%%%%%%%%%%%%%%%%%%

To conclude this discussion of the weak coupling limit let us recall how the
algebraic Bethe ansatz procedure works for the standard $\mathfrak{gl}(1|1)$
rational R-matrix, which, as we have just shown, is relevant in the $h \to 0$
limit. As in section \ref{sec2}, we define $\mathcal T(u)$ as in
\cite{Beisert:2014hya}
\be{
\mathcal T(u)\eqq \sum_{A,B} (-)^{[B]}\,{ e^B}_A \otimes { T^A}_B(u)\ ,
}
with
\ale{
A(u)\eqq { T^1}_1(u) \ , \qquad B(u)\eqq -{ T^2}_1(u)\ ,\nn\\
C(u)\eqq -{ T^1}_2(u)\ ,\qquad D(u)\eqq { T^2}_2(u) \ .
}
The RTT equations determine the commutation relations for $A$, $B$, $C$ and
$D$. In particular, we will need
\ale{
A(\lambda)B(\mu)&=f(\mu,\lambda)B(\mu)A(\lambda)+g(\mu,\lambda)B(\lambda)A(\mu)\ ,\nn\\
D(\lambda)B(\mu)&=h(\lambda,\mu)B(\mu)D(\lambda)+k(\lambda,\mu)B(\lambda)D(\mu)\ ,\nn\\
B(\lambda)B(\mu)&=-q(\lambda,\mu)B(\mu)B(\lambda) \ ,
}
where
\ale{
f(\mu,\lambda)&\eqq R_{11}(\mu,\lambda)/R_{33}(\mu,\lambda)\ ,\nn\\
g(\mu,\lambda)&\eqq R_{23}(\mu,\lambda)/R_{33}(\mu,\lambda)\ ,\nn\\
h(\lambda,\mu)&\eqq R_{44}(\lambda,\mu)/R_{33}(\lambda,\mu)\ ,\nn\\
k(\lambda,\mu)&\eqq -R_{23}(\lambda,\mu)/R_{33}(\lambda,\mu)\ ,\nn\\
q(\lambda,\mu)&\eqq R_{44}(\lambda,\mu)/R_{33}(\lambda,\mu)\ .
}

The standard rational monodromy matrix admits a pseudo-vacuum state $\Omega$,
such that
\be{
A(\lambda)\Omega=\alpha(\lambda)\Omega \ ,\qquad C(\lambda)\Omega=0 \ ,\qquad D(\lambda)\Omega=\delta(\lambda)\Omega \ .
}
This implies that $\Omega$ is an eigenstate of the transfer matrix
\be{
t(\lambda)\eqq (-)^{[I]}{ T^I}_I(u)=A(\lambda)-D(\lambda) \ .
}
The Bethe equations arise from requiring that the $M$-magnon state
\be{
\Phi(\mu_1,\dots,\mu_M)\eqq \prod_{i=1}^MB(\mu_i)\,\Omega \ ,
}
is an eigenstate of $t(\lambda)$. For instance, for $M=2$ one gets
\be{\label{ddd}
\f{\alpha(\mu_1)}{\delta(\mu_1)}=\f{h(\mu_1,\mu_2)\,k(\lambda,\mu_1)}{f(\mu_2,\mu_1)\,g(\mu_1,\lambda)} \ .
}
Substituting in the entries of the rational (weak coupling) R-matrix, we see
that \eqref{ddd} is the same as \eqref{eq: beforffmagn}. This is the expected
result for the $\gl(1|1)$ R-matrix\fn{Let us note that this would actually be
true for both the twisted and the untwisted R-matrix.} (see for instance
\cite{beis1,Kazakov:2007fy}).

%%%%%%%%%%%%%%%%%%%%%%%%%%%%%%%%%%%%%%
\section{Conclusions}\label{comments}
%%%%%%%%%%%%%%%%%%%%%%%%%%%%%%%%%%%%%%

In this paper, we have performed a series of studies on the conjectured exact
S-matrix for the massive excitations of the $AdS_2 \times S^2 \times T^6$
superstring. This S-matrix encodes the integrability of the quantum problem,
and is supposed to be the first step towards the complete solution of the
theory in the ``planar" limit (no joining or splitting of strings). This in
turn is expected to provide information on the spectrum of the elusive
superconformal quantum mechanics, which should be holographically related to
the superstring in this background.

Our main results are as follows:
\begin{itemize}
\item By employing the technique of the RTT realisation, we have found the
presence of Yangian symmetry for the massive sector, and given two alternative
presentations -- both in the spirit of Drinfeld's second realisation \cite{DII}
-- along with the map relating them. We have studied the Yangian coproduct, and
found the conditions under which we can have a consistent evaluation
representation. In order to ascertain these requirements, we studied the
co-commutativity of the higher central charges, which is a necessary condition
for the existence of an R-matrix. We discovered that shortening is one way to
have a consistent evaluation representation, exactly as it was noticed in
$AdS_5$ \cite{Arutyunov:2009pw}. However, we demonstrated explicitly that there
is a second route, which crucially for the $AdS_2 \times S^2$ superstring holds
for long representations.
\item We also found, as in the higher dimensional cases, a secret symmetry,
which is present only at level 1 of the Yangian and higher. This confirms the
ubiquitous presence of this symmetry in all the known manifestations of
integrability in AdS/CFT.
\item We have studied the {\it classical $r$-matrix} of the problem, and
rediscovered from its analytic structure the need for the extra $\mathfrak{gl}$
type (secret) generator. This is also similar to the situation in the $AdS_5$
case.
\item We have taken the first steps towards a derivation of the Bethe equations
starting from our S-matrix (inverse scattering method). At zero coupling, we
discovered that our S-matrix becomes (up to a twist which is easily dealt with)
two copies of the standard rational $\mathfrak{gl}(1|1)$ R-matrix. This allows
one to define a pseudo-vacuum in each copy individually, and enormously
simplifies the problem. Taking the zero-coupling limit of the S-matrix as a
generating R-matrix for the Algebraic Bethe Ansatz, we obtain an effective
model of free fermions on a periodic spin-chain. Let us note that this should
relate to the would-be {\it one-loop} result of the first nested level in the
Bethe ansatz of the putative spin-chain, describing the superconformal quantum
mechanics dual to the superstring.
\end{itemize}

There are a number of future directions which we plan to explore:
\begin{itemize}
\item It seems that we are re-discovering many of the features of the $AdS_5$
(and, to a certain extent, the $AdS_3$) Yangian. In the $AdS_2$ case, however,
the algebra is small enough that we are able to say a more, especially in terms
of alternative presentations. This means that we may hope to find the complete
Drinfeld second realisation and derive the universal R-matrix through a
suitable ansatz. This would also help to understand the higher dimensional
cases. In turn, we would then be able to finally construct the much sought
after exotic quantum group, which should quantise the classical $r$-matrix
algebra, and would prove the algebraic integrability of the system.
\item The most urgent challenge is probably to derive the full set of Bethe
(Beisert-Staudacher) equations for the spectral problem. They encode the
information of the planar anomalous dimensions in the dual theory, and would
hence provide vital information regarding the nature of the holographic dual to
the $AdS_2$ superstring theory. The simplifying assumption of zero coupling of
course eliminates those entries which are responsible for the full
pseudo-vacuum being a mixed state in the two copies. Therefore, a more
sophisticated technique, rather than the simple algebraic Bethe ansatz
computation we have performed here, might be required. This should tie in with
a thorough off-shell worldsheet analysis in the spirit of
\cite{Arutyunov:2006ak,BogdanLatest}.
\item Further directions include studying D-branes in this background, and
performing a boundary integrability analysis as recently done in
\cite{Prinsloo:2014dha}. Also, it would be illuminating to continue the
perturbative and unitarity analyses of \cite{amsw,swnew,pert,uc}, obtaining
further information on the dressing phase and the dispersion relation.
\end{itemize}

%%%%%%%%%%%%%%%%%%%%%%%%%%%%%%%%%%%%%%
\section*{Acknowledgments}
%%%%%%%%%%%%%%%%%%%%%%%%%%%%%%%%%%%%%%

We would like to thank A. Tseytlin for useful discussions. B.H. is supported by the Emmy Noether Programme ``Gauge fields from Strings''
funded by the German Research Foundation (DFG). A.P. is supported in part by
the EPSRC under the grant EP/K503186/1. A.T. thanks the EPSRC for funding under
the First Grant project EP/K014412/1 ``Exotic quantum groups, Lie superalgebras
and integrable systems", and the STFC for support under the Consolidated Grant
project nr. ST/L000490/1 ``Fundamental Implications of Fields, Strings and
Gravity". A.T. also acknowledges useful conversations with the participants of
the ESF and STFC supported workshop ``Permutations and Gauge String duality"
(STFC - 4070083442, Queen Mary U. of London, July 2014).

No data beyond those presented and cited in this work are needed to validate this study.

%%%%%%%%%%%%%%%%%%%%%%%%%%%%%%%%%%%%%%

%%%%%%%%%%%%%%%%%%%%%%%%%%%%%%%%%%%%%%

%%%%%%%%%%%%%%%%%%%%%%%%%%%%%%%%%%%%%%
\end{document}